\documentstyle[epsf,times]{mn}

\title[The ROSAT Brightest Cluster Sample (BCS) --- IV. The extended sample]
      {The ROSAT Brightest Cluster Sample (BCS) --- IV. The extended sample}
\author[H.\ Ebeling et al.]
       {\parbox{\textwidth}{H.\ Ebeling$^{1,2,3}$,    
        A.C.\ Edge$^2$, 
        S.W.\ Allen$^2$, 
        C.S.\ Crawford$^2$, A.C.\ Fabian$^2$, 
        J.P.\ Huchra$^4$ }\\ \\
        $^1$ Max-Planck-Institut f\"ur extraterrestrische Physik, 
             Giessenbachstr., D-85740 Garching, Germany\\
 	$^2$ Institute of Astronomy, Madingley Road, Cambridge CB3\,0HA, UK\\
        $^3$ Institute for Astronomy, 2680 Woodlawn Dr, Honolulu HI 96822, USA;
             email: {\em ebeling{@}ifa.hawaii.edu}\\
	$^4$ Harvard-Smithsonian Center for Astrophysics, 60 Garden Street, 
             Cambridge, MA 02138, USA} 

\date{To appear in MNRAS}
\begin{document}

\maketitle

\begin{abstract} 

We present a low-flux extension of the X-ray selected ROSAT Brightest
Cluster Sample (BCS) published in Paper I of this series. Like the
original BCS and employing an identical selection procedure, the BCS
extension is compiled from ROSAT All-Sky Survey (RASS) data in the
northern hemisphere ($\delta \geq 0^{\circ}$) and at high Galactic
latitudes ($|b| \geq 20^{\circ}$). It comprises 100 X-ray selected
clusters of galaxies with measured redshifts $z \leq 0.3$ (as well as
seven more at $z> 0.3$) and total fluxes between $2.8\times 10^{-12}$
erg cm$^{-2}$ s$^{-1}$ and $4.4\times 10^{-12}$ erg cm$^{-2}$ s$^{-1}$
in the 0.1--2.4 keV band (the latter value being the flux limit of the
original BCS).  The extension can be combined with the main sample
published in 1998 to form the homogeneously selected extended BCS
(eBCS), the largest and statistically best understood cluster sample
to emerge from the ROSAT All-Sky Survey to date.

The nominal completeness of the combined sample (defined with respect
to a power law fit to the bright end of the BCS $\log N-\log S$
distribution) is relatively low at 75 per cent (compared to 90 per
cent for the high-flux sample of Paper I). However, just as for the
original BCS, this incompleteness can be accurately quantified, and
thus statistically corrected for, as a function of X-ray luminosity
and redshift.

In addition to its importance for improved statistical studies of the
properties of clusters in the local Universe, the low-flux extension
of the BCS is also intended to serve as a finding list for X-ray
bright clusters in the northern hemisphere which we hope will prove
useful in the preparation of cluster observations with the next
generation of X-ray telescopes such as {\sl Chandra} or {\sl XMM-Newton}.

An electronic version of the eBCS can be obtained from the following
URL: {\em www.ifa.hawaii.edu/$\sim$ebeling/clusters/BCS.html.}

\end{abstract}


\section{Introduction} 

Until recently, the compilation of large statistical samples of
clusters of galaxies was a task accomplishable only at optical
wavelengths where photographic plates provide both all-sky coverage
and sufficient depth to detect clusters at redshifts of $z\la 0.3$
(e.g., Abell 1958, Zwicky et al.\ 1961--1968, Abell, Corwin \& Olowin
1989). Only with the completion of the {\sl ROSAT} All-Sky Survey
(RASS) in 1991 (Voges 1992, Tr\"umper 1993) did unbiased large
compilations of X-ray detected clusters become a feasible alternative.

To date, three X-ray flux limited cluster samples have been published
from RASS data. The all-sky sample of the 242 X-ray Brightest
Abell-type Clusters (XBACs) of Ebeling et al.\ (1996) was the first
statistical sample of X-ray bright clusters to emerge from the
RASS. However, although X-ray flux limited, the XBACs sample is, by
design, limited to Abell clusters and thus still affected by the
biases inherent in optical cluster surveys. The other two large-scale
RASS cluster samples are truly X-ray selected though: the ROSAT
Brightest Cluster Sample (BCS, Ebeling et al.\ 1998, Paper I)
comprises 203 X-ray selected clusters in the northern hemisphere, and
the RASS1 Bright Sample (RASS1-BS, De Grandi et al.\ 1999) consists of
130 such clusters in the southern hemisphere. A fourth RASS cluster
sample covering most of the southern extragalactic sky is under
compilation (B\"ohringer et al, in preparation).  In the following we
briefly summarize the key features of the BCS.

\section{The ROSAT Brightest Cluster Sample (BCS)} 

The BCS is the first truly X-ray selected, and so far largest, cluster
sample to emerge from the RASS.  The BCS, as listed in Paper I,
comprises 203 RASS selected galaxy clusters in the northern hemisphere
($\delta \geq 0^{\circ}$) and at high Galactic latitudes ($|b| \geq
20^{\circ}$). All 203 BCS clusters have measured redshifts of $z\le
0.374$; the 201 BCS clusters at $z\le 0.3$ form the statistical
subsample which is nominally 90\% flux complete.

\begin{figure}
  \epsfxsize=0.5\textwidth
  \hspace*{-0.5cm} \centerline{\epsffile{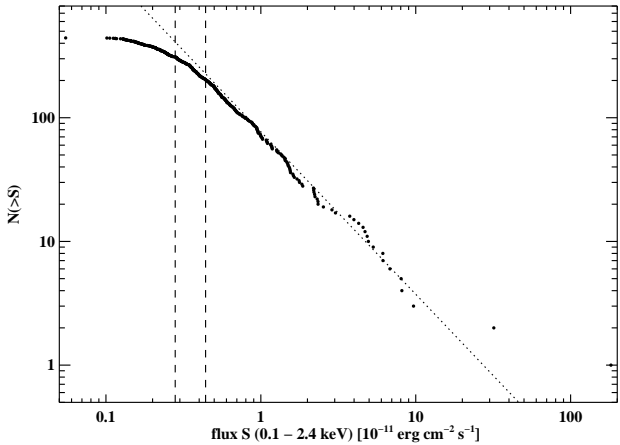}}
  \caption[]{The cumulative flux distribution of all 442 clusters in
           our sample of VTP detections with raw count rates greater
           than 0.07 count s$^{-1}$. The dotted line represents the
           power law description of the BCS $\log N-\log S$
           distribution as determined in Paper I. The dashed lines
           mark the flux limits of the original BCS and the extended
           sample (eBCS) defined by the requirement of 90 and 75 per cent
	   completeness, respectively.}  \label{flux_lim}
\end{figure}

The primary selection criterion for the BCS is X-ray extent as
measured in the RASS with the Standard Analysis Software System (SASS)
used in the original analysis of the raw data. Since clusters of
galaxies are, even at the moderate resolution of the RASS, generically
extended X-ray sources at essentially all redshifts, this approach
should reliably select all clusters\footnote{A small number of other
extended X-ray sources, such as supernova remnants, can be eliminated
easily}.  Unfortunately, the SASS source detection algorithm is known
to erroneously misclassify a significant fraction of clusters as point
sources. As demonstrated in the appendix of Paper I, a RASS cluster
sample based exclusively on SASS source extent will miss 25\% of all
clusters at all redshifts, nearly independent of the imposed X-ray
flux limit. `Pure' X-ray selection by source extent is thus bound to
result in significant incompleteness. In the compilation of the BCS we
overcome this limitation of the SASS by adding to the clusters found
through their SASS extent all X-ray detections (irrespective of
alleged X-ray extent) of Abell and Zwicky clusters. We also perform a
comprehensive reanalysis of the raw RASS photon data around all SASS
detected clusters (the total sky area thus processed amounts to one
sixth of the BCS survey area) with VTP (Voronoi Tesselation and
Percolation, Ebeling \& Wiedenmann 1993), a source detection algorithm
developed and optimized specifically for the detection and
characterization of irregular, extended X-ray emission from galaxy
clusters.  This step is crucial in two respects. Firstly, it yields
accurate total cluster fluxes and, secondly, it ensures that the BCS,
contrary to other RASS cluster projects, does not exclusively rely on
detections made by a point source detection algorithm. It is the
additional clusters detected in this second processing that allow us
to quantify and correct for the incompleteness and bias introduced by
the imperfections of the initial point source detection algorithm.

As a result of this strategy, details of which can be found in Paper
I, the BCS is currently not only the largest but, perhaps more
importantly, also the statistically best understood RASS cluster
sample.  The BCS has been used for a wide range of astrophysical
studies, including investigations of the spectral properties of
central cluster galaxies (Allen et al.\ 1992, Crawford et al.\ 1995,
1999), the currently best determination of the cluster X-ray
luminosity function at $z\le 0.3$ (Ebeling et al.\ 1997) and the
bright end of the cluster X-ray $\log N-\log S$ distribution (Ebeling
et al.\ 1998). The BCS is also used routinely in studies of cluster
evolution to define properties of the cluster population in the local
universe (e.g., Kitayama, Sasaki \& Suto 1998, Rosati et al.\ 1998,
Jones et al.\ 1998, Blanchard \& Bartlett 1998, Vikhlinin et al.\
1999, Reichert et al.\ 1999, Nichol et al.\ 1999, Ebeling et al.\
2000).

Work in progress investigates the cosmological implications of the BCS
X-ray luminosity function in the framework of a Press-Schechter model
(Ebeling et al., in preparation) and the three-dimensional large-scale
distribution of clusters via the cluster-cluster correlation function
(Edge et al., in preparation).  Both of these studies use the extended
BCS as defined and described below.

As in Paper I, we assume an Einstein-de Sitter Universe with $q_0 =
0.5$ and $H_0 = 50$ km s$^{-1}$ Mpc$^{-1}$ throughout.

\section{The BCS extension} 

The flux limit of the BCS extension is defined crudely by the
requirement that the supplementary sample comprise about 100 clusters
below the flux limit of the main sample released in 1998\footnote{An
agreement with MPE requires the authors to limit the total size of the
BCS as published to about 300 clusters.}. Following the same procedure
as for the original BCS we find this requirement to be met for a flux
limit of the extended BCS (eBCS) of $2.8\times 10^{-12}$ erg cm$^{-2}$
s$^{-1}$ (0.1--2.4 keV), corresponding to a flux completeness of 75
per cent. The correspondence between completeness and X-ray flux limit
is illustrated in Figure~\ref{flux_lim} which, similar to Fig.~20 of
Paper~I, shows the observed BCS $\log N-\log S$ distribution, as well
as the power law fit to the same distribution after corrections for
incompleteness have been applied (see Paper I for details). The eBCS
flux limit is defined by the X-ray flux (note that these are total
cluster fluxes) at which the observed distribution falls below 3/4 of
the value predicted by the power law description of the cumulative
flux distribution for a complete sample.

The BCS extension thus defined comprises 107 clusters from the total
sample of 442 clusters found in the RASS data during the compilation
of the BCS. All 107 have measured redshifts of $z\le 0.418$; 100 of
them fall within the redshift limit of the complete sample at $z\le
0.3$. By design, the X-ray fluxes of the clusters in the extension
range from $2.8\times 10^{-12}$ erg cm$^{-2}$ s$^{-1}$ to $4.4\times
10^{-12}$ erg cm$^{-2}$ s$^{-1}$ in the 0.1--2.4 keV band (the latter
value being the flux limit of the original BCS).

Table~1 lists the 107 clusters in the BCS extension in
analogy to Table~3 of Paper I, i.e., the contents of this table are
\begin{list}{}{\labelwidth17mm \leftmargin17mm}
  \item[column\,\,\, 1:] redshift, contamination, extent, and serendipity flag.
		   Clusters at $z>0.3$ are marked $\star$; `c' means a 
		   significant fraction of the quoted flux may come from 
                   embedded point sources; `V' (`S') signals significant X-ray
                   extent according to VTP (SASS), i.e., an extent value in 
                   excess of 35 arcsec; systems flagged by a $\bullet$ symbol 
                   are serendipitous VTP detections in the sense of 
 		   Section~7.1 of Paper I.
  \item[column\,\,\, 2:] cluster name. Where clusters appear to consist of two 
		   components, two entries (`a' and `b') are listed.
                   We adopt cluster names in the following order of
                   priority: Abell name, Zwicky name, other cluster name
                   established in the literature, ROSAT RXJ name.
  \item[column\,\,\, 3:] right ascension (J2000) of the X-ray position as
                   determined by VTP.
  \item[column\,\,\, 4:] declination (J2000) of the X-ray position as 
                   determined by VTP.
  \item[column\,\,\, 5:] column density of Galactic Hydrogen from Stark et al.
                   (1992).
  \item[column\,\,\, 6:] total RASS exposure time.
  \item[column\,\,\, 7:] PSPC count rate in PHA channels 11 to 235 originally
                   detected by VTP.
  \item[column\,\,\, 8:] equivalent radius $\sqrt{A_{\rm VTP}/\pi}$ of
                   the source detected by VTP. 
  \item[column\,\,\, 9:] final PSPC count rate in Pulse Height Analyzer (PHA) 
                   channels 11 to 235
                   based on the original VTP count rate. Statistical 
		   corrections for low-surface
                   brightness emission that has not been detected directly and
                   for contamination from point sources have been applied.
  \item[column 10:] error in the final PSPC count rate according to 
		   equation~4 of Paper I. The fractional uncertainty in the 
		   energy flux (column 13) and the X-ray luminosity (column 14)
 		   can be assumed to be the same as the fractional count rate 
		   error.
  \item[column 11:] ICM gas temperature used in the conversion from count
 		   rates to energy fluxes. `e' indicates the temperature has
		   been estimated from the $L_{\rm X}-{\rm k}T$ relation.
  \item[column 12:] measured redshift. 
  \item[column 13:] unabsorbed X-ray energy flux in the 0.1 to 2.4 keV band.
  \item[column 14:] intrinsic X-ray luminosity in the 0.1 to 2.4 keV band
		   (cluster rest frame).
  \item[column 15:] reference for the redshift in column 12.
\end{list}

One of the clusters listed in Table~1, A1758a, was already listed in
Table~3 of Paper I because it made the flux limit of the original
sample when combined with its X-ray fainter companion A1758b. The latter
falls below the flux limit of both the original and the extended BCS.

\begin{figure} 
  \epsfxsize=0.5\textwidth
  \hspace{-0.5cm} \centerline{\epsffile{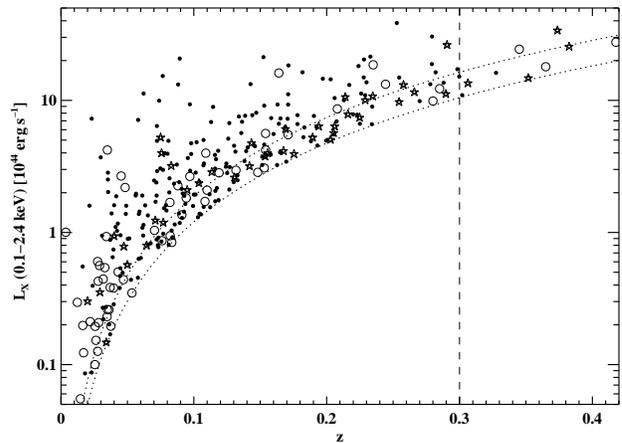}}
  \caption{The X-ray luminosity of the 310 clusters of the 75 per cent
	    complete flux limited eBCS as a function of redshift. The
	    dotted lines show the cutoff introduced by the X-ray flux
	    limits at $2.8 \times 10^{-12}$ erg cm$^{-2}$ s$^{-1}$ (75
	    per cent completeness) and $4.4 \times 10^{-12}$ erg
	    cm$^{-2}$ s$^{-1}$ (90 per cent completeness, Paper I).
	    Abell (Zwicky) clusters are plotted as solid dots (stars);
	    the remaining 57 clusters not contained in these largest
	    optical cluster catalogues are shown as open circles.}
	    \label{ebcs_lum_z}
\end{figure}

The distribution of the full eBCS sample of 310 clusters in
luminosity-redshift space is shown in Fig.~\ref{ebcs_lum_z}. While 211
(68\%) of these are Abell clusters, the number of non-Abell clusters
in the eBCS is substantial: 42 (14\%) systems are Zwicky clusters
without Abell identification, and another 57 (18\%) are listed in
neither of the two largest optical cluster catalogues. As expected, the
Abell content of the eBCS is thus somewhat lower than, but still
similar to, the one found for the BCS where the fractional content in
Abell, Zwicky and other clusters was measured to be 70, 11, and 19 per
cent.

\begin{figure} 
  \epsfxsize=0.5\textwidth
  \hspace*{-0.5cm} \centerline{\epsffile{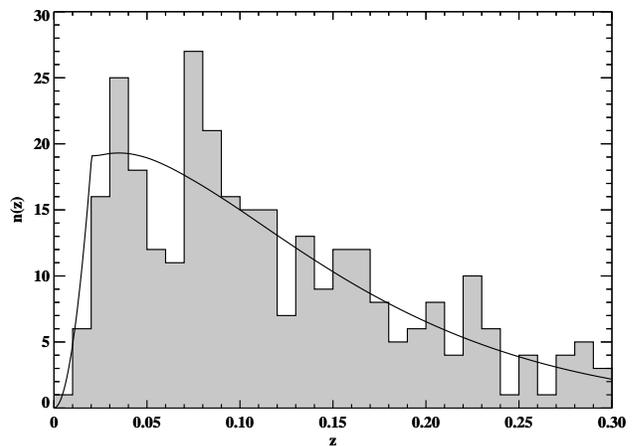}}
  \caption{The differential redshift distribution of the clusters in
            the extended BCS. The solid curve shows the redshift
            distribution expected for a spatially homogeneous cluster
            distribution.  }  \label{ebcs_n_z}
\end{figure}

The redshift distribution of the 310 clusters of the extended BCS
shows striking signs of large-scale structure as already noted in the
original BCS. Figure~\ref{ebcs_n_z} shows the eBCS redshift histogram
compared to the distribution expected for a spatially homogenous
sample. The pronounced peaks at $z\sim 0.036$ and $z\sim 0.077$, as
well as the depletion between them, are prominent also in the redshift
distribution of the original BCS (see Section 8.1 of Paper I). As
shown in Paper I, the excess of clusters at these redshifts can not be
attributed to any single supercluster but is generated by clusters and
superclusters distributed widely over the solid angle covered by the
BCS.

\begin{figure*} 
  \hspace*{-2cm}
  \parbox{0.55\textwidth}{
  \epsfxsize=0.54\textwidth
  \epsffile{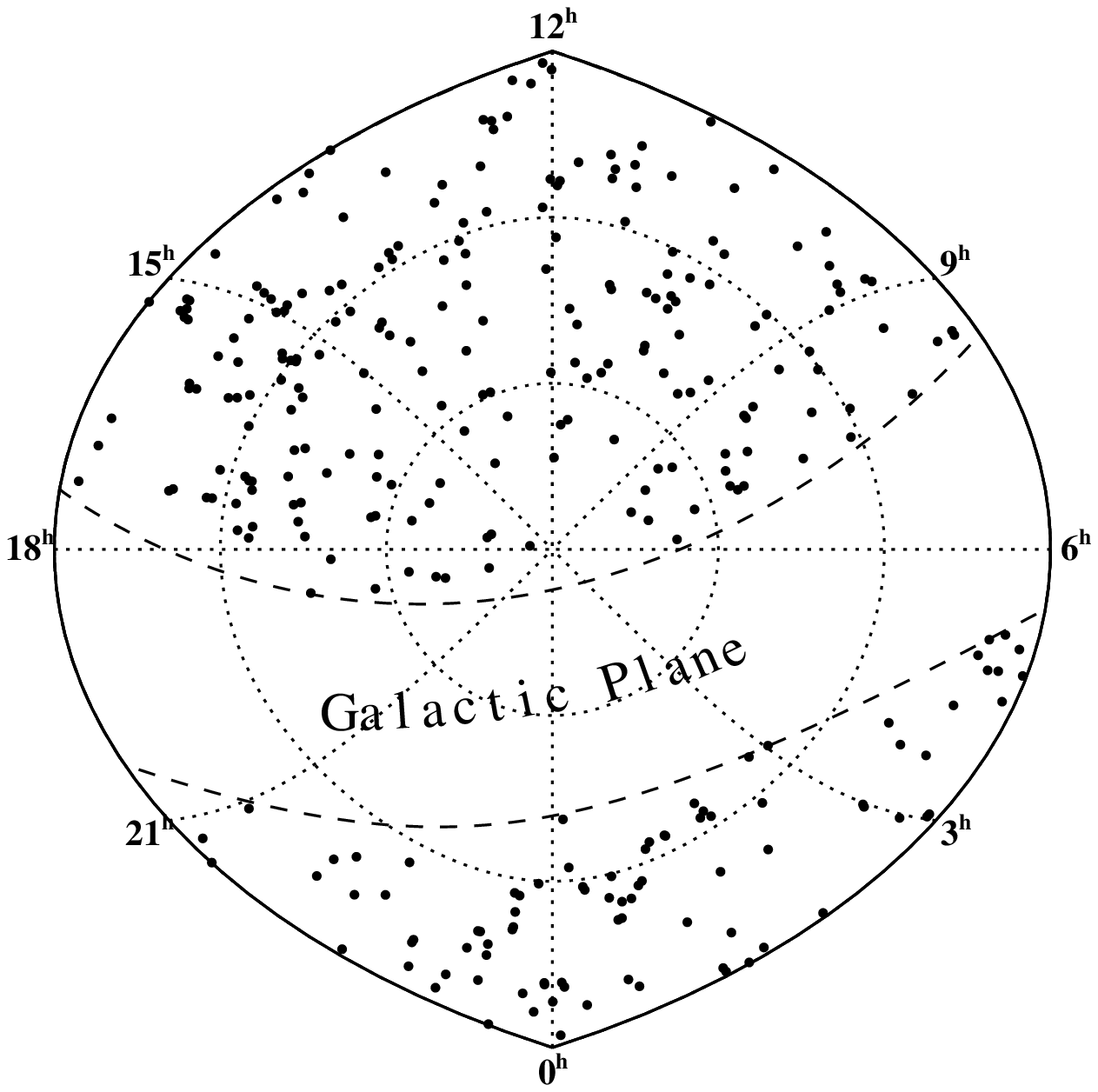}}
  \hspace*{-1cm}
  \parbox{0.55\textwidth}{
  \epsfxsize=0.54\textwidth
  \epsffile{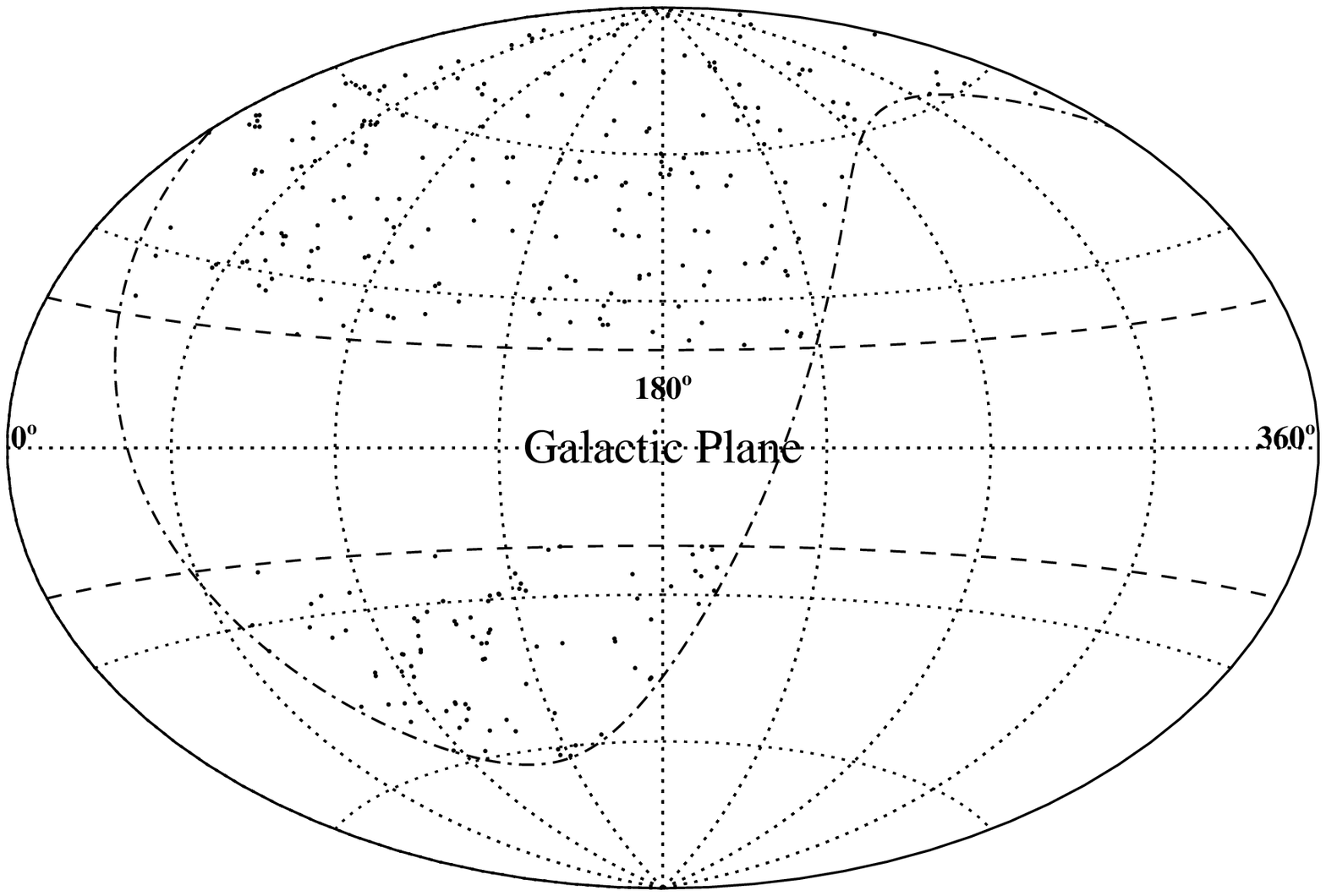}}
  \caption{The distribution of the 301 eBCS clusters ($z\le 0.3$ only)
           on the sky in a
           view of the northern equatorial hemisphere (left) and in
           Galactic coordinates (right).  Equal-area Aitoff
           projections are used in both plots.  The 40 deg wide
           exclusion zone around the Galactic equator is marked by the
           dashed lines. Also marked as a dot-dashed line in the right
           panel is the line of $\delta = 0^{\circ}$.}
           \label{ebcs_skymap}
\end{figure*}

An overview of the distribution of the eBCS on the sky is shown in
Fig.~\ref{ebcs_skymap}. 

\section*{Acknowledgements} 

The authors are indebted to the ROSAT team at MPE for providing the
RASS photon data this analysis is based upon. The identification of
non-cluster sources that contaminated the sample we originally started
from was greatly facilitated by the availability of digitized optical
images from the POSS and UK Schmidt sky surveys obtained through
STScI's DSS Web interface. All of the data analysis presented in this
paper was carried out using the Interactive Data Language (IDL). We
are greatly indebted to all who contributed to the various IDL User's
Libraries; the routines of the IDL Astronomy User's Library
(maintained by Wayne Landsman) have been used particularly
extensively.

HE thanks the Time Allocation Committee of the Institute for Astronomy
for their support of the spectroscopic follow-up observations of eBCS
clusters. HE also gratefully acknowledges partial financial support
from a European Union EARA Fellowship, SAO contract SV4-64008, and
NASA LTSA grant NAG 5-8253. ACE, ACF and CSC thank the Royal Society
for support.

This research has made use of data obtained through the High Energy
Astrophysics Science Archive Research Center Online Service, provided
by the NASA-Goddard Space Flight Center, and the NASA/IPAC
Extragalactic Database (NED).


\begin{thebibliography}{9}
\bibitem{abell} Abell G.O., 1958, ApJS, 3, 211
\bibitem{aco} Abell G.O., Corwin H.G., Olowin R.P., 1989, ApJS, 70, 1
\bibitem{swa} Allen S.W.\ et al., 1992, MNRAS, 259, 67
\bibitem{batuski} Batuski  D.J., Burns J.O., Newberry M.V., Hill J.M.,
 Deeg H.-J., Laubscher Bryan.E., Elston  R.J., 1991, AJ, 101, 1983
\bibitem{beers}  Beers T.C., Kriessler J.R., Bird C.M., Huchra J.P. 1995, AJ, 
	109, 874
\bibitem{bb} Blanchard A.\ \& Bartlett J.G., 1998, A\&A, 332, L49
\bibitem{ciardullo} Ciardullo R., Ford H., Bartko F., Harms R. 1983, ApJ, 273,
 	24
\bibitem{car1} Crawford C.S., Edge A.C., Fabian A.C., Allen S.W., 
	B\"ohringer H., Ebeling H., McMahon R.G., Voges W., 1995, MNRAS, 274, 75
\bibitem{car2} Crawford C.S., Allen S.W., Ebeling H., Edge A.C., Fabian A.C.,
 	1999, MNRAS, 306, 857
\bibitem{sab} De Grandi S.\ et al., 1999, ApJ, 514, 148
\bibitem{ebe0} Ebeling H.\ \& Wiedenmann G., 1993, Phys.~Rev.~E, 47, 704
\bibitem{ebe1} Ebeling H., Voges W., B\"ohringer H., Edge A.C., Huchra J.P.,
	Briel U.G., 1996, MNRAS, 281, 799
\bibitem{ebe2} Ebeling H., Edge A.C., B\"ohringer H., Allen S.W.,
	Crawford C.S., Fabian A.C., Voges W., Huchra J.P., 1998, 
	MNRAS, 301, 881 (Paper I)
\bibitem{ebe3} Ebeling H., Jones L.R., Fairley B., Scharf C., Perlman E., 
        Horner D., 2000, ApJ, submitted
\bibitem{feti} Fetisova T. 1982, Sov.\ Astron., 25, 647
\bibitem{henry} Henry J.P.\ \& Lavery R.J., 1984, ApJ, 280, 1
\bibitem{hoessel} Hoessel J.G., Gunn J.E., Thuan T.X. 1980, ApJ, 241, 466
\bibitem{hopp}  Hopp U., Kuhn, B., Thiele U., Birkle K, Els\"asser H., Kovachev B.,
        1995, A\&AS, 109, 537
\bibitem{hhpg} Huchra J.P., Henry J.P., Postman M., Geller M.J., 1990, ApJ, 
	365, 66
\bibitem{zcat} Huchra J.P., Geller M.J., Clemens C.M., Tokarz S.P., Michel A.
	1992, {\it Bull. C.D.S.}, 41, 31
\bibitem{jones} Jones L.R., Scharf C., Ebeling H., Perlman E., 
	Wegner G., Malkan M., Horner D., 1998, ApJ, 495, 100
\bibitem{kss} Kitayama T., Sasaki S., Suto Y., 1998, PASJ, 50, 1
\bibitem{ledlow} Ledlow M.J. \& Owen F.N., 1995, AJ, 110, 1959
\bibitem{marzke} Marzke R.O., Huchra J.P., Geller M.J., 1996, AJ, 112, 1803
\bibitem{nichol} Nichol R.C.\ et al., 1999, ApJ, 521, L21
\bibitem{owen} Owen F.N., Ledlow M.J., Keel W.C., 1995, AJ, 109, 14
\bibitem{phg} Postman M., Huchra J.P., Geller M.J. 1992, ApJ, 384, 404
\bibitem{postlau} Postman M.\ \& Lauer T.R. 1995, ApJ, 440, 28
\bibitem{rei}  Reichart D.E., Nichol R.C., Castander F.J., Burke D.J.,
   	Romer A.K., Holden B.P., Collins C.A., Ulmer M.P., 1999, ApJ, 518, 521
\bibitem{rhee} Rhee G.\  \& Katgert P. 1988, A\&AS, 72, 243
\bibitem{kath} Romer A.K., 1994, PhD thesis, University of Edinburgh
\bibitem{piero} Rosati P., Della Ceca R., Norman C., Giacconi R., 1998,
	ApJ, 492, L21
\bibitem{sarazin} Sarazin C.L., Rood H.J., Struble M.F. 1982, A\&A, 108, L7
\bibitem{schneider} Schneider D.P., Gunn J.E., Hoessel J.G. 1983, ApJ, 264, 337
\bibitem{shectman} Shectman S.A. 1985, ApJS, 57, 77
\bibitem{small} Small T.A., Sargent W.L.W., Hamilton D., 1997, ApJ, 487, 512
\bibitem{stark} Stark A.A., Gammie C.F., Wilson R.W., Bally J., Linke R.A.,
		  Heiles C., Hurwitz M., 1992 ApJS, 79, 77
\bibitem{stocke} Stocke J.T., Morris S.L., Gioia I.M., Maccacaro T., Schild R.,
		 Wolter A., Fleming T.A., Henry J.P. 1991, ApJS, 76, 813
\bibitem{strubro}  Struble M.F.\ \& Rood H.J. 1987, ApJS, 63, 543
\bibitem{truemper} Tr\"umper J., 1993, Science, 260, 1769
\bibitem{rc3} de Vaucouleurs G., de Vaucouleurs A., Corwin H.G., Buta R.J., 
	Paturel G., Fouqu\'e P. 1991, Third Reference Catalogue of Bright 
	Galaxies, Springer, New York
\bibitem{alex} Vikhlinin A., McNamara B.R., Forman W., Jones C., Quintana H.,
	Hornstrup A., 1999, ApJ, 491, L21
\bibitem{voges} Voges W. 1992, Proceedings of Satellite Symposium 3, ESA ISY-3, p9
\bibitem{annZ} Zabludoff A.I., Huchra J.P., Geller M.J. 1990, ApJS, 74, 1
\bibitem{zwicky} Zwicky F., Herzog E., Wild P., Karpowicz M., Kowal, C.T.,
	1961--1968, {\it Catalogue of galaxies and cluster galaxies}, 
	Vols.\ 1--6
\end{thebibliography}
\end{document}